\documentclass[11pt,3p,times]{elsarticle}
\newenvironment{myproof}{\begin{proof}}{\end{proof}}
\usepackage{amsfonts,amssymb,amscd,amsthm}
\usepackage{latexsym, graphicx}

\newcommand{\La}{\ensuremath{\llcorner}} %L
\newcommand{\Lb}{\ensuremath{\ulcorner}} %\Gamma
\newcommand{\Lc}{\ensuremath{\lrcorner}} %flipped L
\newcommand{\Ld}{\ensuremath{\urcorner}} %flipped \Gamma

\newcommand{\proofBox}{\hfill $\qedsymbol$}

\newtheorem{theorem}{Theorem}
\newtheorem{observation}[theorem]{Observation}
\newtheorem{lemma}[theorem]{Lemma}

\newtheorem{definition}{Definition}
\newtheorem{conjecture}{Conjecture}

\graphicspath{{figures/}}

\journal{Discrete Applied Mathematics: Special Issue for LAGOS'13}

\begin{document}

\author[author1]{Kathie Cameron}
\author[author2]{Steven Chaplick\corref{cor}}
\author[author3]{Ch\'inh T. Ho\`ang}	

\address[author1]{Department of Mathematics, Wilfrid Laurier University, Waterloo, ON, Canada. Email: {\tt kcameron@wlu.ca}}

\address[author2]{Institut f\"ur Mathematik, Technische Universit\"at Berlin, Berlin, Germany. Email: {\tt chaplick@math.tu-berlin.de}}

\address[author3]{Department of Physics and Computer Science, Wilfrid Laurier University, Waterloo, ON, Canada. Email: {\tt choang@wlu.ca} }

\cortext[cor]{Corresponding author.  Phone: {\tt +49 30 314 28706}. Fax: {\tt +49 30 314 25191}}

\begin{frontmatter}

\title{Edge Intersection Graphs of $L$-Shaped Paths in Grids\footnote{An extended abstract of this paper appeared at LAGOS'13 \cite{Cameron2013363}}}

%\author{%
%Kathie Cameron\inst{1}
%Steven Chaplick\inst{2}\thanks{}
%Ch\'inh T. Ho\`ang\inst{3}
%}
%
%\institute{%
%Department of Mathematics, Wilfrid Laurier University, Waterloo,
%ON,  Canada, N2L 3C5, 
%\and
%%Department of Physics and Computer Science, Wilfrid Laurier
%%University, Waterloo, ON, Canada, N2L 3C5, e-mail: {\tt
%%chaplick@cs.toronto.edu}
%Institut f\"ur Mathematik, Technische Universit\"at Berlin,
%Strasse des 17. Juni 136, D-10623 Berlin, Germany,
%email: {\tt chaplick@math.tu-berlin.de}
%\and Department of Physics and Computer
%Science, Wilfrid Laurier University, Waterloo, ON, Canada, N2L
%3C5, e-mail: {\tt choang@wlu.ca} }
%
%\maketitle
%\begin{center}\large\today\end{center}

\begin{abstract}
In this paper we continue the study of the edge intersection graphs of one (or zero) bend paths on a rectangular grid. That is, the edge intersection graphs where each vertex is represented by one of the following shapes: $\La,\Lb,\Lc,\Ld$, and we consider zero bend paths (i.e., $|$ and --) to be degenerate $\La$'s. 
These graphs, called $B_1$-EPG graphs, were first introduced by Golumbic et al (2009). 
We consider the natural subclasses of $B_1$-EPG formed by the subsets of the four single bend shapes (i.e., $\{\La\}, \{\La,\Lb\}, \{\La,\Ld\},$ and $\{\La,\Lb,\Ld\}$) and we denote the classes by $[\La], [\La,\Lb], [\La,\Ld],$ and $[\La,\Lb,\Ld]$ respectively. Note: all other subsets are isomorphic to these up to 90 degree rotation.
We show that testing for membership in each of these classes is NP-complete and observe the expected strict inclusions and incomparability (i.e., $[\La] \subsetneq [\La,\Lb],[\La,\Ld] \subsetneq [\La,\Lb,\Ld] \subsetneq B_1$-EPG
and $[\La,\Lb]$ is incomparable with $[\La,\Ld]$). Additionally, we give characterizations and polytime recognition algorithms for special subclasses of \textit{Split}~$\cap$~$[\La]$. 
\end{abstract}


\end{frontmatter}

\section{Introduction}
A graph $G$ is called an {\em EPG graph} if $G$ is  the
intersection graph of paths on a rectilinear grid, where each vertex in $G$
corresponds to a path on the grid and two vertices are adjacent in
$G$ if and only if the corresponding paths share an edge on the grid. EPG
graphs were introduced by Golumbic et al \cite{Gol2009}. The
motivation for studying these graphs comes from circuit layout
problems \cite{Ban1990}. Golumbic et al \cite{Gol2009} defined a {\em
$B_k$-EPG graph} to be the edge intersection graph of paths on a
grid where the paths are allowed to have at most $k$ bends
(turns). The $B_0$-EPG graphs are exactly the well studied {\em
interval graphs} (the intersection graphs of intervals on a line).

Golumbic and Jamison \cite{Gol1985}
proved that  the recognition problem for the edge intersection
graphs of paths in trees (EPT) is NP-complete even when restricted
to chordal graphs (i.e., graphs without induced $k$-cycles for $k \geq 4$). Heldt et al \cite{Hel2010} proved that the recognition problem for
$B_1$-EPG is NP-complete.
In a recent paper Epstein et al \cite{EpsteinGM13} have demonstrated that both the
coloring problem and the independent set problem are NP-complete on $B_1$-EPG graphs.
They have further shown that these problems can be 4-approximated in polynomial time
when a $B_1$-EPG representation is given and that the clique problem can be solved
optimally in polynomial time even without a given EPG representation.

A graph is {\em chordal} if it does not
contain a chordless cycle with at least four vertices as an
induced subgraph. A graph is a {\em split graph} if its vertices
can be partitioned into a clique and a stable set; {\em Split}
denotes the class of split graphs. Asinowski and
Ries \cite{Asi2012} characterized special subclasses of chordal
$B_1$-EPG graphs.

Consider a $B_1$-EPG graph $G$ with a path representation on a grid.
The paths can be of the following four shapes: $\La, \Lb, \Lc,
\Ld$. In this paper, we  study  $B_1$-EPG graphs whose paths on
the grid belong to a proper subset of the four shapes.   
If ${\cal S}$
is  a subset of $\{\La, \Lb, \Lc, \Ld\}$, then $[{\cal S}]$
denotes the class of graphs that can be represented by paths whose
shapes belong to ${\cal S}$. It is important to note that we also 
zero-bend paths (i.e., vertical and horizontal line segments) to be $\La$s. 
In particular, an $[\La]$-representation of a graph may have some of its vertices represented as zero-bend paths. 
We are especially interested in the class
$[\La]$ of $B_1$-EPG graphs whose paths are of the type $\La$. Our
main results are:
\begin{itemize}
 \item Establishment of expected separation between the classes: $[\La] \subsetneq
[\La,\Lb],[\La,\Ld] \subsetneq [\La,\Lb,\Ld] \subsetneq B_1$-EPG
and the incomparability between $[\La,\Lb]$ and $[\La,\Ld]$.
 \item A proof of NP-completeness of recognition of $[\La]$ and
 of each of the other subclasses of $B_1$-EPG mentioned above.
 \item Characterizations of, and recognition algorithms for
 gem-free split $[\La]$-graphs and bull-free split $[\La]$-graphs.
\end{itemize}
In Section~\ref{sec:properties}, we discuss background results and
establish some properties of $B_1$-EPG graphs. In
Section~\ref{sec:np-hard}, we show that recognition of $[\La]$
and of each of the other subclasses is
an NP-complete problem. In Section~\ref{sec:split}, we give
polytime recognition algorithms for the classes of
gem-free split $[\La]$-graphs and of bull-free split $[\La]$-graphs. We
conclude with some open questions in Section~\ref{sec:conclusion}.
%
%\section{Preliminaries}
\section{Properties of $B_1$-EPG graphs}\label{sec:properties}
\begin{figure}[h]
\hfill
\includegraphics[scale=1.5]{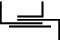}
\hfill
\includegraphics[scale=1.5]{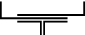}
\hfill \ \caption{Left: An edge-clique. Right: A
claw-clique.}\label{fig:edgeclique}
\end{figure}

Let $\mathcal{P}$ be a collection of nontrivial simple paths on a rectilinear
grid $\mathcal{G}$. (The end-points of each path are grid points.)
The \textit{edge intersection graph}
$EPG(\mathcal{P})$ has a vertex $v$ for each path $P_v \in
\mathcal{P}$ and two vertices are adjacent in $EPG(\mathcal{P})$
if the corresponding paths in $\mathcal{P}$ share an edge of
$\mathcal{G}$. For any grid edge $e$, the set of paths containing
$e$ is a clique in $EPG(\mathcal{P})$; such a clique is called an
\textit{edge-clique} \cite{Gol2009}. A \textit{claw} in a grid
consists of three grid edges meeting at a grid point.  The set of
paths which contain two of the three edges of a claw is a clique;
such a clique is called a \textit{claw-clique} \cite{Gol2009} (see
Figure~\ref{fig:edgeclique}).
\begin{lemma}[\cite{Gol2009}]\label{lem:Gol2009}
Consider a $B_1$-EPG representation
%on a grid
of a graph $G$.
Every clique in $G$ corresponds to either an edge-clique or a claw-clique.
\end{lemma}
%
%Golumbic et al \cite{Gol2009} proved that in a $B_1$-EPG
%representation of a graph $G$, every clique in $G$ is either an
%edge-clique or a claw-clique.

The \textit{neighborhood} $N(x)$ of a vertex $x$ is the set of vertices adjacent to $x$.
A set of vertices is \textit{stable} if no two are adjacent. An \textit{asteroidal triple}
(AT) is a stable set of size three such
that for every pair, there is a path between them which avoids the
neighborhood of the other vertex.

\begin{lemma}[AT Lemma \cite{Asi2012}, Theorem 9]\label{lem:AT}
In a $B_1$-EPG graph, no vertex can have an AT in its
neighborhood.
\end{lemma}

Let $C_4$ denote the chordless cycle $a,b,c,d,a$ on four vertices.
Golumbic et al \cite{Gol2009} proved that any $B_1$-EPG
representation of $C_4$ corresponds to what they call a ``true
pie", a ``false pie", or a ``frame".  True and false pies require
paths other than $\La$'s. A \textit{frame} is a rectangle in the
grid $\mathcal{G}$ such that each corner is the bend-point for one
of $P_a,P_b,P_c$ and $P_d$;  $P_a \cap P_b,P_b \cap P_c,P_c \cap
P_d$, and $P_d \cap P_a$ each contain at least one grid edge; and  $P_a
\cap P_c$ and $P_b \cap P_d$ each do not contain an grid edge. Consider
the $C_4$ and four representations of it shown in
Figure~\ref{fig:C4-K23}. The first three representations are frames,
the fourth is a false pie, and the fifth is a true pie. It follows
that:

\begin{figure}[bth]
\hfill
\includegraphics[scale=1]{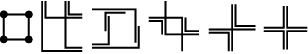}
\hfill
\includegraphics[scale=1]{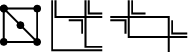}
\hfill \ \caption{Left: $C_4$ and representations of it. Right:
$K_{2,3}$ and representations of it.}\label{fig:C4-K23}
\end{figure}
\begin{lemma}[$C_4$ Lemma]\label{lem:C4}
In an $[\La]$- or $[\La,\Lb]$-representation of a $C_4$, every $\La$,  and $\Lb$ has a neighbor on both its vertical segment and on its horizontal segment.
\end{lemma}

\begin{observation}
$K_{2,3}$ is in $[\La,\Ld]$.
\end{observation}
\begin{myproof}
%Proof.
See Figure~\ref{fig:C4-K23} for an $[\La,\Ld]$-representation of $K_{2,3}$.
\end{myproof}

\begin{lemma}[$K_{2,3}$ Lemma]\label{lem:K2,3}
In an $[\La,\Ld]$-representation of a $K_{2,3}$ every $\La$, and $\Ld$ has a neighbor on both its vertical segment and on its horizontal segment.
\end{lemma}
\begin{myproof}
Consider $K_{2,3}$
to be the complete bipartite graph with bipartition $\{\{a,b\},$
$\{c,d,e\}\}$. Note that each of the following is a $C_4$:
$a,c,b,d,a$; $a,c,b,e,a$; and $a,d,b,e,a$. As noted above, any
$B_1$-EPG representation of $C_4$ corresponds to  a ``true pie", a
``false pie", or a ``frame".  True pies require paths of all four
types, but false pies and frames can be made from just $\La$'s and
$\Ld$'s.

%If an $[\La,\Ld]$-representation of a $C_4$ corresponds to a
%frame, then every $\La$ (and $\Ld$) has a neighbor on both
%its vertical segment and on its horizontal segment. Consider an
%$[\La,\Ld]$-representation of a $K_{2,3}$.

%Suppose first that both  $\{a,c,b,d\}$ and $\{a,c,b,e\}$
%correspond to frames.  Then $P_d$ and $P_e$ must have the same
%bend-point, and this bend-point must be an  intersection point of
%$P_a$ and $P_b$. Since $d$ and $e$ are not adjacent, one of $P_d$
%and $P_e$ is an $\La$ and the other is an $\Ld$. It follows that
%every $\La$ (and $\Ld$) has a neighbor on both its vertical
%segment and on its horizontal segment.

%Now suppose that $\{a,c,b,d\}$ corresponds to a false pie. If
%$P_a$ and $P_b$ have the same bend-point, the bend-point must be
%an intersection point of $P_c$ and $P_d$, and then there is
%nowhere to place $P_e$ so that it intersects both $P_a$ and $P_b$.
%So it must be that $P_c$ and $P_d$ have the same bend-point, which
%must be an intersection point of $P_a$ and $P_b$, say point $p$.
%Then $P_e$ must have bend-point at an intersection point of $P_a$
%and $P_b$, but since $e$ is not adjacent to $c$ or to $d$, this
%must be a different intersection point from $p$. So we have a
%configuration such as that in Figure~\ref{fig:C4-K23}, and it
%follows that every $\La$ (and $\Ld$) has a neighbor on both its
%vertical segment and on its horizontal segment. (Note that
%$\{a,c,b,e\}$ corresponds to a frame.)

%It is not possible for both $\{a,c,b,d\}$ and $\{a,c,b,e\}$ to
%both correspond to false pies.
%\end{myproof}

If an $[\La,\Ld]$-representation of a $C_4$ corresponds to a
frame, then every $\La$ (and $\Ld$) has a neighbor on both
its vertical segment and on its horizontal segment. Consider an
$[\La,\Ld]$-representation of a $K_{2,3}$. Either at least two of the $C_4$'s
correspond to frames or at least two of the $C_4$'s correspond to false pies.
The latter is clearly not possible.

Suppose that both  $\{a,c,b,d\}$ and $\{a,c,b,e\}$
correspond to frames.  Then $P_d$ and $P_e$ must have the same
bend-point, and this bend-point must be an  intersection point of
$P_a$ and $P_b$. Since $d$ and $e$ are not adjacent, one of $P_d$
and $P_e$ is an $\La$ and the other is a $\Ld$. It follows that
every $\La$ (and $\Ld$) has a neighbor on both its vertical
segment and on its horizontal segment. Note that $\{a,d,b,e\}$ corresponds
to a false pie.
\end{myproof}
\begin{observation}\label{obs:K2,3}
$K_{2,3}$ is in $[\La,\Ld]$ but not in $[\La,\Lb]$. 
\end{observation}
\begin{myproof}
Again, recall
that $a,c,b,d,a$ and $a,c,b,e,a$ are $C_4$'s in $K_{2,3}$. True and
false pies are not representable using just $\La$'s and $\Lb$'s.  So
both of these must be represented as frames.  As argued above,
$P_d$ and $P_e$ must have the same bend-point.  But since $d$ and
$e$ are not adjacent, if $P_d$ is an $\La$, then $P_e$ must be an
$\Ld$ and vice versa. It follows that $K_{2,3}$ is not in
$[\La,\Lb]$.
\end{myproof}
\begin{figure}[bth]
\hfill
\includegraphics[scale=1.2]{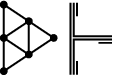}
\hfill
\includegraphics[scale=1.5]{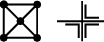}
 \hfill \ \caption{Left: 3-sun and a
representation of it. Right: 4-wheel and a
representation of it.}\label{fig:3-sun-and4-wheel}
\end{figure}
\begin{observation}\label{obs:3-sun}
The 3-Sun is in $[\La,\Lb]$ but not in $[\La,\Ld]$.
\end{observation}
\begin{myproof}
See Figure~\ref{fig:3-sun-and4-wheel} for the 3-sun and an
$[\La,\Lb]$-representation of it. To see that the 3-sun
does not have an $[\La,\Ld]$-representation, recall that in a
$B_1$-EPG graph, every clique is an edge-clique or a claw-clique.
The vertices of the 3-sun can be partitioned into a clique with
vertices $a,b,c$ and a stable set with vertices $d,e,f$ with edges
$da, dc, ea, eb, fb, fc$.  It is easy to see that if the clique
$\{a,b,c\}$ is an edge-clique, then only two of $d,e,f$ can be
represented regardless of which types of 1-bend paths are used. So
the clique $\{a,b,c\}$ is a claw-clique. But $\La$'s and $\Ld$'s
can not form a claw-clique.
\end{myproof}
\begin{observation}\label{obs:4-wheel}
The 4-wheel is in $[\La,\Lb,\Ld]$ but not in $[\La,\Lb]$ or
$[\La,\Ld]$.
\end{observation}
\begin{myproof}
See Figure~\ref{fig:3-sun-and4-wheel} for the 4-wheel $W_4$ and an
$[\La,\Lb, \Ld]$-representation of it. Lemma 3 in
\cite{Asi2012} shows that in a $B_1$-representation of $W_4$, the
$C_4$ corresponds to a true pie or a false pie. Since the true pie
requires four shapes, we may assume the $C_4$ of the $W_4$ is
represented by a false pie. So, $W_4$ is not an $[\La,\Lb]$-graph.
Consider the vertex $u$ of $W_4$ that is adjacent to all vertices
of the $C_4$. If $P_u$ is of type $\La$ or $\Ld$, then $P_u$ can
not share a grid edge with all four paths of the $C_4$. So, the
$W_4$ is not an $[\La,\Ld]$-graph.
\end{myproof}

\begin{figure}[bth]
\centering
\includegraphics[scale=1.5]{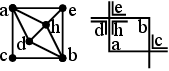}
\caption{$Y_6$ and an $[\La,\Lb,\Ld]$ representation of it.}
\label{fig:Y6}
\end{figure}

Let $Y_6$ denote the graph shown in Figure
\ref{fig:Y6}.
Graph $Y_6$ consists of $K_{2,3}$ with bipartition $\{\{a,b\},\{c,d,e\}\}$
together with a vertex $h$ adjacent to all the other vertices except $c$. Note
that $Y_6$ contains both $K_{2,3}$ and $W_4$ as induced subgraphs, and thus is
not representable in $[\La,\Lb]$ or $[\La,\Ld]$. Figure
\ref{fig:Y6}
gives an $[\La,\Lb, \Ld]$-representation of $Y_6$.

\begin{lemma}[$Y_6$ Lemma]\label{lem:Y6}
In any $[\La,\Lb,\Ld]$-representation of $Y_6$ every $\La$, $\Lb$, and $\Lc$ has a neighbor on both its vertical segment and on its horizontal segment.
\end{lemma}
\begin{myproof}
As mentioned above, Lemma 3 in
\cite{Asi2012} implies that in an $[\La,\Lb,\Ld]$-representation of $W_4$, the
$C_4$ is represented by a false pie. Consider an $[\La,\Lb,\Ld]$-representation of $Y_6$.
The induced $W_4$ of $Y_6$ is represented as in Figure
\ref{fig:Y6Proof}
(i). The $\La$ and $\Ld$ of Figure 
\ref{fig:Y6Proof}
(i) are either $P_a$ and $P_b$ or $P_d$ and $P_e$. Vertex $c$ of $Y_6$ is adjacent to vertices $a$ and $b$ only.
It follows that the $\La$ and $\Ld$ of Figure
\ref{fig:Y6Proof}
(i) are $P_d$ and $P_e$, and that $P_a$ and $P_b$ intersect in a second point $Q$, which is the bend-point of $P_h$
(see Figure
\ref{fig:Y6Proof}
(ii) ). The representation is unique up to whether $P_a$ is an $\La$ and $P_b$ is a $\Ld$ or vise versa,
and the shape of $P_h$ (see Figure
\ref{fig:Y6Proof}
(iii) for example). In any case, each $\La$, $\Lb$ and $\Ld$ has a neighbor on its vertical segment
and on its horizontal segment.

\begin{figure}[h]
\hfill
\begin{tabular}{c}
\includegraphics[scale=1.5]{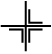}\\
(i)
\end{tabular}
\hfill
\begin{tabular}{c}
\includegraphics[scale=1.5]{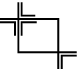}\\
(ii)
\end{tabular}
\hfill
\begin{tabular}{c}
\includegraphics[scale=1.5]{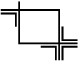}\\
(iii)
\end{tabular}
\hfill \
\caption{Visual aids for the proof of Lemma \ref{lem:Y6}.}
\label{fig:Y6Proof}
\end{figure}
\end{myproof}

\section{NP-Hardness: Recognition of $[\La]$ and of Other Subclasses of $B_1$-EPG }
\label{sec:np-hard}

It is well-known that interval graphs (i.e., $B_0$-EPG graphs) can
be recognized in polynomial time \cite{Boo1976}. The complexity of
the recognition problem for $B_k$-EPG ($k > 0$) was given as an
open problem in the paper introducing EPG graphs \cite{Gol2009}.
The recognition problem for $B_1$-EPG has been shown to be
NP-complete in a recent paper \cite{Hel2010}. In this section we
consider the complexity of recognizing the simplest natural
subclass of $B_1$-EPG which is a superclass of $B_0$-EPG; namely,
$[\La]$. Specifically, we show that it is NP-complete to decide
membership in $[\La]$. We will call the classes $[\La],
[\La,\Lb],[\La,\Ld]$ and $[\La,\Lb,\Ld]$ \textit{the natural
subclasses of} $B_1$-EPG. We show that it is NP-complete to
decide membership in each of these classes.

We use $G[A]$ to denote the subgraph of $G$ induced by the set $A$
of vertices.

\begin{theorem}
Deciding membership in each of $[\La],
[\La,\Lb],[\La,\Ld]$ and $[\La,\Lb,\Ld]$  is NP-complete.
\end{theorem}
\begin{myproof}
A given $[\La]$ model is easily verified, so $[\La]$ recognition
is in NP, and the same is true for each of the other classes.
For NP-hardness we demonstrate a reduction from the
usual 3-SAT problem (defined below). Our reduction is inspired by
the NP-completeness proof for $B_1$-EPG \cite{Hel2010}.

The essential ingredients of our construction are described in the
following observations. In a $B_1$-EPG-representation $R$ of a
graph $G$ containing vertices $u$ and $v$, we say that $v$ is an
\emph{internal neighbor} of $u$ in $R$ when: $v$ is adjacent to
$u$, $P_u$'s bend-point is not contained in $P_v$ and w.l.o.g.
$P_u$'s horizontal contains $P_v$'s horizontal (see Figure
\ref{fig:internal-external}(i)). We also say that $v$ is an
\emph{external neighbor} of $u$ when $v$ is adjacent to $u$ but
$v$ is not an internal neighbor of $u$. Notice that, in any
$B_1$-EPG-representation of a graph, a vertex can have at most four
stable external neighbors (as depicted in Figure
\ref{fig:internal-external}(ii)). Additionally, if a vertex $v$ is
an internal neighbor of a vertex $u$, then $v$ can have at most
two stable external neighbors which are not adjacent to $u$ (see
Figure \ref{fig:internal-external}(iii)). Finally, we say that a
vertex $u$ is \emph{adjacent to a graph $H$} when $u$ is adjacent to
exactly one vertex in an induced $H$ (see Figure
\ref{fig:internal-external}(iv) where $H=C_4$).

Let $\mathcal{N}=\{[\La],[\La,\Lb],[\La,\Ld],[\La,\Lb,\Ld]\}$ denote
the set of natural subclasses of $B_1$-EPG.  We will use $\mathcal{B}$ to denote
an arbitrary natural subclass.  For each natural subclass
$\mathcal{B} \in \mathcal{N}$, we define a special graph $F(\mathcal{B})$:
$F([\La])=F([\La,\Lb])=C_4.$ $F([\La,\Ld])=K_{2,3}.$ $F([\La,\Lb,\Ld])=Y_6.$

Recall that for each $\mathcal{B} \in \mathcal{N}$, in any
$\mathcal{B}$-representation of $F(\mathcal{B})$, every $\La$, $\Lb$ and
$\Ld$ of the representation has a neighbor with edge-intersection on
its vertical and a neighbor with edge-intersection on its horizontal (by
Lemmas \ref{lem:C4}, \ref{lem:K2,3} and \ref{lem:Y6}).

Consider a graph $G \in \mathcal{B}$ with a vertex $u$ that is adjacent to an
$F(\mathcal{B})$, and let $v$ be $u$'s neighbor in $F(\mathcal{B})$. It follows
from the previous paragraph that in any $\mathcal{B}$-representation of $G$, $v$ is
necessarily an external neighbor of $u$.

With these observations
in mind we can now describe the structure of our graph $G_{\Phi,\mathcal{B}}$.
\begin{figure}[h]
\begin{tabular}{c}
\includegraphics[scale=1]{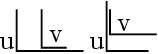} \\
(i)
\end{tabular}
\hfill
\begin{tabular}{c}
\includegraphics[scale=1]{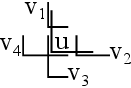}\\
(ii)
\end{tabular}
\hfill
\begin{tabular}{c}
\includegraphics[scale=1]{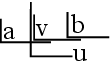}\\
(iii)
\end{tabular}
\hfill
\begin{tabular}{c}
\includegraphics[scale=1]{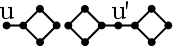}\\
(iv)
\end{tabular}
\caption{(i): $v$ is an internal neighbor of $u$ (left:
\emph{internal horizontal  neighbor}; right: \emph{internal
vertical neighbor}). (ii): $u$ with stable external neighbors
$v_1,v_2,v_3,v_4$. (iii): $v$ is an internal neighbor of $u$, and
$v$ has two stable external neighbors $a,b$ which are not adjacent
to $u$. (iv): $u$ adjacent to one $C_4$ and $u'$ adjacent to two
adjacent $C_4$'s.} \label{fig:internal-external}
\end{figure}
A \emph{3-SAT} formula $\Phi$ is a boolean formula over variables
$x_1, ..., x_k$ where $\Phi$ is a conjunction of $t$ clauses
$D_1, D_2, ..., D_t$, each clause $D_i$ ($1\leq i \leq t$) is a
disjunction of three literals $\ell_{i1}, \ell_{i2}, \ell_{i3}$,
and each literal $\ell_{iq}$ ($1\leq q \leq 3$) is either
%the negation or non-negation of
some variable $x_j$ ($1 \leq j \leq k$)
or its negation.
Given a 3-SAT formula $\Phi$, it is well known that it is
NP-complete to decide whether there exists an assignment to the
variables of $\Phi$ that satisfies $\Phi$ \cite{Kar1972}.

Given a 3-SAT formula $\Phi$ and $\mathcal{B} \in \mathcal{N}$,
we will construct a graph $G_{\Phi,\mathcal{B}}$
such that $G_{\Phi,\mathcal{B}}$ is in $\mathcal{B}$ if and only if $\Phi$ can be
satisfied. Graph $G_{\Phi,\mathcal{B}}$ consists of an induced subgraph $G_{D_i}$ for
each clause $D_i$ of $\Phi$ and a variable gadget to identify the
clauses with their corresponding literals. The general form of
these gadgets where $\mathcal{B}=[\La]$ or $\mathcal{B}=[\La,\Lb]$ and thus
$F(\mathcal{B})=C_4$ is given in Figure \ref{fig:gadgets}.
\begin{figure}[h]
\centering
\hfill
\includegraphics[scale=1]{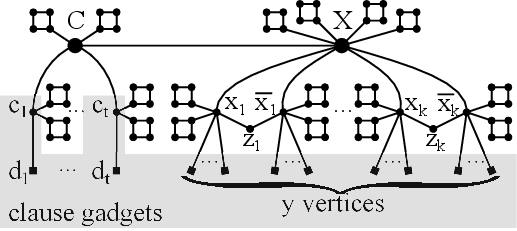}
\hfill
\includegraphics[scale=1]{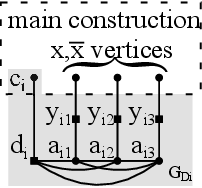}
\hfill \ 
\caption{The general form of $G_{\Phi,\mathcal{B}}$ when $\mathcal{B}=[\La]$ or
$\mathcal{B}=[\La,\Lb]$. On the left is the main construction of $G_{\Phi,\mathcal{B}}$ where the clause gadgets (depicted on the right) are drawn in the shaded region. Also, the shaded region of the depiction of the single clause gadget (on the right) corresponds to the induced subgraph $G_{D_i}$ of $G_{\Phi,\mathcal{B}}$ and the dotted box contains the remainder of $G_{\Phi,\mathcal{B}}$. Note: the literals (i.e., vertices of the form $x_j$ or
$\overline{x_j}$) of the ith clause ($D_i$) are matched to $y_{i1},y_{i2},y_{i3}$ in the clause gadget
$G_{D_i}$.} \label{fig:gadgets}
\end{figure}

For $\mathcal{B}=[\La,\Ld]$, each of the $C_4$'s that $C$, $X$, and the
$c_i$'s are adjacent to is replaced by $K_{2,3}$ (so that $C$, $X$, and the
$c_i$'s are adjacent to a degree 2 vertex of $K_{2,3}$; also, for each j,
$1 \leq j \leq k$, we add a vertex $z'_j$ which is adjacent to
$x_j$ and $\overline{x_j}$ (thus turning the $C_4$ induced by
$\{X, x_j, z_j, \overline{x_j}\}$ into a $K_{2,3}$.

Similarly, for $\mathcal{B}=[\La,\Lb,\Ld]$, each of the $C_4$'s that $C$, $X$, and the
$c_i$'s are adjacent to is replaced by $Y_6$ (so that $C$, $X$, and the
$c_i$'s are adjacent to the degree 2 vertex of $Y_6$); also, for each j,
$1 \leq j \leq k$, we add a vertex $z'_j$ which is adjacent to
$x_j$ and $\overline{x_j}$ as above and a vertex $z''_j$ adjacent to
$x_j, \overline{x_j}, z_j$ and $z'_j$  (thus turning the $C_4$ induced by
$\{X, x_j, z_j, \overline{x_j}\}$ into a $Y_6$).

We begin by describing the structure of the $\mathcal{B}$-representation
of the variable gadget. Notice that the vertex $X$ is adjacent to
four $F(\mathcal{B})$'s. Thus, as we have observed, $X$ will have four
external neighbors in any $\mathcal{B}$-representation of $G_{\Phi,\mathcal{B}}$.
Furthermore, since the neighborhood of $X$ is a stable set, the
vertices $C, x_1, x_2, ..., x_k, \overline{x_1}, \overline{x_2},
..., \overline{x_k}$ are all internal neighbors of $X$. Without loss of
generality, we will assume that $X$ is represented by an $\La$. Finally,
suppose that $x_j$ is an internal horizontal neighbor of $X$. When $\mathcal{B}$
is $[\La]$ or $[\La,\Lb]$,
since $G_{\Phi,\mathcal{B}}[\{X,x_j,z_j, \overline{x_j}\}]$ is a $C_4$,
which can only be represented as a frame,
$\overline{x_j}$ is necessarily an internal vertical neighbor of
$X$. For $\mathcal{B}=[\La,\Ld]$,
$G_{\Phi,\mathcal{B}}[\{X,x_j,z_j, \overline{x_j},z'_j\}]$ is $K_{2,3}$,
with bipartition $\{X,z_j,z'_j\}$ and $\{x_j,\overline{x_j}\}$. In an
$[\La,\Ld]$-representation of $K_{2,3}$, only a vertex of the size 2 partite set
can have two internal horizontal neighbors.  So $\overline{x_j}$ is necessarily an
internal vertical neighbor of $X$. For $\mathcal{B}=[\La,\Lb,\Ld]$,
$G_{\Phi,\mathcal{B}}[\{X,x_j,z_j, \overline{x_j},z'_j, z''_j\}]$ is $Y_6$ where
$X$ is the vertex of degree 2.  In an $[\La,\Lb,\Ld]$-representation of $Y_6$, only
the neighbors of the degree 2 vertex can have two internal horizontal neighbors.
So again, $\overline{x_j}$ is necessarily an
internal vertical neighbor of $X$. Similarly, if $x_j$ were to be an internal vertical neighbor
of $X$, $\overline{x_j}$ would necessarily be an internal
horizontal neighbor of $X$ (\footnote{We will later use the
location (i.e., as an internal horizontal or internal vertical
neighbor of $X$) as a variable's truth value.}).

From these
observations, in Figure \ref{fig:var-gadget}, for $\mathcal{B}=[\La]$,
we depict the general structure of a
$[\La]$-representation of the subgraph of $G_{\Phi,[\La]}$ induced by
$\{X,$ $x_1,$ $...,$ $x_k,$ $\overline{x_1},$ $...,$
$\overline{x_k},$ $z_1,$ $...,$ $z_k\}$ and the $C_4$'s adjacent to
these vertices.

Now, w.l.o.g., suppose that $C$ is an internal horizontal neighbor
of $X$. Notice that $C$ is adjacent to two $F(\mathcal{B})$'s, that $C$ is an internal
horizontal neighbor of $X$, and that the neighborhoods of $X$ and $C$
are disjoint. Thus, since the neighborhood of $C$ is a stable set,
the vertices $c_1, ..., c_t$ are internal vertical neighbors of
$C$. Similarly, for each $1\leq i \leq t$, $d_i$ is an internal
horizontal neighbor of $c_i$ since each $c_i$ is an internal
vertical neighbor of $C$ and each $c_i$ is adjacent to two $F(\mathcal{B})$'s.
These observations provide the general structure of a
$\mathcal{B}$-representation of the subgraph of $G_{\Phi,\mathcal{B}}$ induced by
$\{X,$ $C,$ $c_1,$ $...,$ $c_t,$ $d_1,$ $...,$ $d_t\}$ and the
$F(\mathcal{B})$'s adjacent to these vertices (as seen in Figure
\ref{fig:var-gadget} for $\mathcal{B}=[\La]$).
\begin{figure}[h]
\hfill
\includegraphics[scale=1]{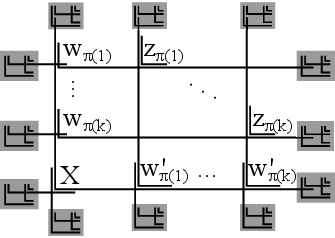}
\hfill
\includegraphics[scale=1]{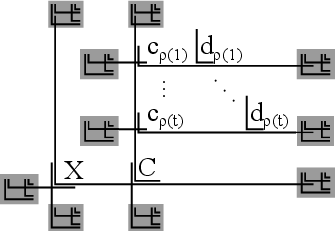}
\hfill \ 
\caption{Left: The possible $[\La]$-representations of $G_{\Phi,[\La]}$
induced by $\{X,$ $x_1,$ $...,$ $x_k,$ $\overline{x_1},$ $...,$
$\overline{x_k},$ $z_1,$ $...,$ $z_k\}$ and the $C_4$'s adjacent to
these vertices (note: $w_i \in \{x_i, \overline{x_i}\}$ and
$\{w_i, w'_i\} = \{x_i, \overline{x_i}\}$, and $\pi$ is a
permutation on $\{1, ..., k\}$). Right: The possible
$[\La]$-representations of $G_{\Phi,[\La]}$ induced by $\{X,$ $C,$ $c_1,$
$...,$ $c_t,$ $d_1,$ $...,$ $d_t\}$ and the $C_4$'s adjacent to
these vertices (note: $\rho$ is a permutation on $\{1, ...,
t\}$).} \label{fig:var-gadget}
\end{figure}

With the restricted structure of the variable gadget in mind, we
now turn our attention to the clause gadget of a clause $D_i =
\ell_{i1} \vee \ell_{i2} \vee \ell_{i3}$. Notice that $\{d_i, a_{i1}, a_{i2}, a_{i3}\}$ is
a clique (i.e., $\{P_{d_i},P_{a_{i1}},P_{a_{i2}},P_{a_{i3}}\}$ have pairwise edge-intersections
in any $\mathcal{B}$-representation of $G_{\Phi,\mathcal{B}}$).
Furthermore, $P_{a_{i1}}$, $P_{a_{i2}}$ and $P_{a_{i3}}$ intersect
$P_{d_i}$'s vertical only since $d_i$ is an internal horizontal
neighbor of $c_i$. Only the vertical with the highest top-point
and the vertical with the lowest bottom-point are not contained
in the union of the other three verticals.  W.l.o.g., assume that the
vertical of $P_{a_{i2}}$ is contained in the union of the verticals of
$P_{d_i}$, $P_{a_{i1}}$ and $P_{a_{i3}}$. Since $d_i$, $a_{i1}$
and $a_{i3}$ are not adjacent to $y_{i2}$, paths $P_{a_{i2}}$ and
$P_{y_{i2}}$ must not intersect in a vertical grid edge, but rather in a
horizontal grid edge (see Figure \ref{fig:clause-gadget}).
Additionally, observe that, when
$\ell_{iq} (1 \leq q \leq 3)~$\footnote{Remember, $\ell_{iq} (1 \leq q \leq 3)$ is some $x_j$ or
$\overline{x_j}$ ($1\leq j \leq k$).} is an internal vertical
neighbor of $X$, $y_{iq}$ is necessarily an internal horizontal
neighbor of $\ell_{iq}$ since $\ell_{iq}$ is adjacent to two $F(\mathcal{B})$'s.
Similarly, when $\ell_{iq}$ is an internal horizontal neighbor of
$X$, $y_{iq}$ is an internal vertical neighbor of $\ell_{iq}$. However,
$y_{i2}$ cannot be an internal horizontal neighbor of $\ell_{i2}$ since
$\ell_{i2}$ is not adjacent to $a_{i2}$ and $P_{y_{i2}}$ and $P_{a_{i2}}$ have
a horizontal grid edge in common. Thus, it is not possible for all
three literals to be internal vertical neighbors of $X$. On the
other hand, when at most two literals are internal vertical
neighbors of $X$, we can always construct the
$\mathcal{B}$-representation of the clause gadget. In particular, this
can be done using one of the three templates depicted in Figure
\ref{fig:clause-gadget}. Note, to form an $\mathcal{B}$-representation
of $G_{\Phi,\mathcal{B}}$, the placement of the $\mathcal{B}$-representations of the
clause gadgets from Figure \ref{fig:clause-gadget} can be described as
follows:
\begin{itemize}
\item When at most one literal is an
internal vertical neighbor of $X$, (i.e. for type (i) and (ii) of
Figure \ref{fig:var-gadget}), we place the
$\mathcal{B}$-representation of the clause gadget ``below''
$P_{w_{\pi(k)}}$ and to the ``left'' of $P_{w'_{\pi(1)}}$ (with
respect to the depiction in Figure \ref{fig:var-gadget}).
\item When two literals $\ell_{i1}$ and $\ell_{i3}$ are
internal horizontal neighbors of $X$, (i.e. for type (iii)
of Figure \ref{fig:var-gadget}), we need to place the
$\mathcal{B}$-representation of the clause gadget ``between''
$P_{\ell_{i1}}$ and $P_{\ell_{i3}}$ and to the ``left'' of
$P_{w'_{\pi(1)}}$ (with respect to the depiction in Figure
\ref{fig:var-gadget}).
\end{itemize}
\begin{figure}[h]
\begin{tabular}{c}
\includegraphics[scale=1]{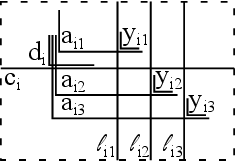} \\
(i)
\end{tabular}
\hfill
\begin{tabular}{c}
\includegraphics[scale=1]{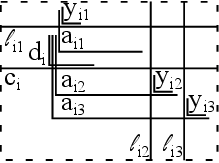} \\
(ii)
\end{tabular}
\hfill
\begin{tabular}{c}
\includegraphics[scale=1]{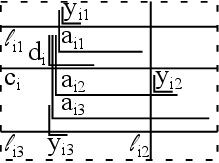} \\
(iii)
\end{tabular}
\caption{$[\La]$-representations of the clause gadget for a clause
$\ell_{i1} \vee \ell_{i2} \vee \ell_{i3}$ inside an $[\La]$-representation of
$G_{\Phi,[\La]}$. (i) $\ell_{i1} = \ell_{i2} = \ell_{i3} = true$; (ii) $\ell_{i1} =
false$ and $\ell_{i2} = \ell_{i3} = true$; (iii) $\ell_{i1} = \ell_{i3} =
false$ and $\ell_{i2} = true$.} \label{fig:clause-gadget}
\end{figure}
We can now see that a literal being an internal vertical neighbor
of $X$ corresponds to when that literal is \emph{false} (since at
most two literals can be internal vertical neighbors of $X$) and a
literal being an internal horizontal neighbor of $X$ corresponds
to when that literal is \emph{true}. Thus, since $x_j$ and
$\overline{x_j}$ cannot both be internal vertical (or horizontal)
neighbors of $X$, the $\mathcal{B}$-representations of 
$G_{\Phi,\mathcal{B}}$ correspond to satisfying assignments of $\Phi$.
\end{myproof}

%An interesting observation regarding this proof is that it can be
%easily adapted to show the NP-completeness of recognition for
%$[\La,\Lb]$ and $[\La,\Ld]$. For $[\La,\Lb]$, the same graph
%$G_\Phi$ can be used, but one has to be careful about the
%structure of the $[\La,\Lb]$-representation of the clause gadget
%(since it need not be an edge clique). For $[\La,\Ld]$, we alter
%$G_\Phi$ slightly. First, we replace each $C_4$ adjacent to a
%vertex with a $K_{2,3}$ adjacent to the same vertex (by Lemma
%\ref{lem:K2,3}, this forces the vertex to have an external
%neighbor for each $K_{2,3}$, just as we had with the $C_4$'s in
%$[\La]$-representations). Second, for each $1 \leq i \leq k$, we
%add a vertex $z'_i$ which is adjacent to $x_i$ and
%$\overline{x_i}$ (thus, turning the $C_4$'s induced by
%$\{X,x_i,\overline{x_i},z_i\}$ into $K_{2,3}$'s and preventing
%$x_i$ and $\overline{x_i}$ from both being internal vertical
%(horizontal) neighbors of $X$). These two changes to $G_\Phi$
%allow the proof to proceed as before.

\noindent We conjecture that a similar approach can be used to prove
that recognizing $B_k$-EPG is NP-hard for $k>1$. 
%the NP-completeness of recognizing $B_k$-EPG for $k>1$.

\section{Characterization and Recognition of {\em Split} $\cap$ $[\La]$}
\label{sec:split}
Recall that recognizing chordal EPT graphs is NP-complete \cite{Gol1985}. We have
just shown that recognizing $[\La]$-graphs is NP-complete. Thus,
it is of interest to characterize the class {\em Chordal} $ \cap
\; [\La]$. A first step in this direction would be to study {\em
Split} $ \cap \; [\La]$, that is, the class of split
$[\La]$-graphs. We divide this discussion into three parts. In the
first part, we establish some properties of split $[\La]$-graphs.
In the latter two parts, we characterize two special subclasses of
split $[\La]$-graphs.
\subsection{Properties of {\em Split} $\cap$ $[\La]$}
\label{sec:properties-of-split}
In this section, we will establish some properties of the class
{\em Split} $ \cap \; [\La]$. We conjecture a
characterization of this class. First, we need to introduce a few
definitions.

Recall that $N(x)$ denotes the set of vertices adjacent to vertex $x$.
Vertices $x$ and $y$ are called {\em twins} if either they are non-adjacent
and $N(x)=N(y)$ or if they are adjacent and $N(x) \cup \{x\} = N(y) \cup \{y\}$.
A vertex $x$ {\em dominates} a vertex $y$ if $N(y) \subseteq N(x)
\cup \{x\}$. The domination relation is reflexive and transitive, but need not be
antisymmetric - twins dominate each other.   Two vertices are
{\em comparable} if one dominates the other. A vertex is called {\em maximal}
if it is not dominated by any other vertex.

Let $X$ be a subset of vertices of $G=(V,E)$. A vertex which belongs
to $X$ is called an {\em $X$-vertex}. We use $N(X)$ to denote the set of
vertices not in $X$ which have at least one neighbor in $X$.  We use $G-X$  to denote the
subgraph of $G$ induced by the vertices of $G$ which are not in $X$.
%Two vertices $a,b$ are {\em twins} if $N(a) \cup \{a\} = N(b) \cup \{b\}$ or $N(a) = N(b)$.

We say that an $\La$-path lies on a horizonal (vertical) line $Q$ if its horizontal (vertical) part
intersects $Q$ in a grid edge. An $\La$-path $L_1$ lies on another $\La$-path $L_2$ if part of $L_1$
lies on part of $L_2$. We say an $\La$-path $L_1$ lies above (below) another $\La$-path or horizontal
line $L_2$ if the y-coordinate of the horizontal part of $L_1$ is greater (less) than the y-coordinate
of the horizontal part of $L_2$. Lying to the left or right is defined similarly.

A {\em split partition} $(C,S)$ of a graph $G$ is a partition of
its vertices into a clique $C$ and a stable set $S$. We will
enumerate the vertices of $S$ as $\{s_1, ..., s_k\}$.

\begin{figure}[h]
\hfill
\includegraphics[scale=1]{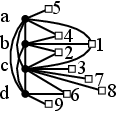}
\hfill
\includegraphics[scale=1]{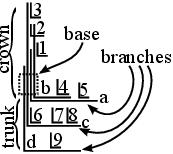}
\hfill \ \caption{A Split $\cap [\La]$ graph (left) and an
$[\La]$-representation of it (right).}\label{fig:L-representation}
\end{figure}

Let $G$ be an $[\La]$-graph  with a split partition $(C,S)$. Consider an
$[\La]$-representation of $G$ on the grid. It
follows from Lemma \ref{lem:Gol2009} that $C$ corresponds to an
edge-clique. We may assume without loss of generality that the edge of
the grid that belongs to all $\La$-paths of $C$ is vertical. The horizontal
parts of $\La$-paths of $C$ are called {\em branches}. Let $F$ be the vertical
line-segment which is the union of the vertical parts of all $\La$-paths of $C$.
The part of $F$ below the first (top) branch is called the {\em trunk}. The
part of $F$ above the first branch is called the {\em
crown} (see Figure~\ref{fig:L-representation}). All $\La$-paths of $C$ contain the
lowest grid-edge of the crown; call this the {\em base} of the crown.

%When we say ``placing a vertex $x$ on a branch (trunk, crown)'' we
%mean ''placing the path $P_x$ on a branch (trunk, crown)''.

%{\bf For the remainder of this section, $G$ is assumed to be a
%split graph with a split partition $(C,S)$} (ie., not necessarily
%an $[\La]$-graph.)
%

\begin{observation}\label{obs:comparable}
The $S$-vertices whose $\La$-paths lie on the same branch (or
on the crown) are pairwise comparable. An $S$-vertex whose $\La$-path lies
on the trunk dominates all $S$-vertices whose $\La$-paths lie below it
in the representation. \proofBox
\end{observation}
See Figure~\ref{fig:L-representation} for an illustration of
Observation~\ref{obs:comparable}.

The {\em gem} is the graph with vertices $a,b,c,d,e$, edges
$ab,bc,cd,ea,eb,ec,ed$. The {\em bull} is the graph with vertices
$a,b,c,d,e$, edges $ab,bc,cd,eb,ec$; vertex $e$ is called the {\em nose}
of the bull.  In a split partition $(C,S)$ of the path $P_4$ on four
vertices, the degree 1 vertices are in $S$ and the degree 2 vertices
are in $C$. It follows that any split partition of the gem has $a$ and $d$
in $S$ and $b$, $c$ and $e$ in $C$.  In a split partition of the bull,
$a$ and $d$ are in $S$ and $b$ and $c$ are in $C$, but the
nose $e$ may be in either $C$ or $S$. When the nose is in $S$, the
bull is called an {\em $S$-bull}; that is,

\begin{definition}\label{def:S-bull}
An {\em S-bull} is a bull such that the three vertices of degrees
less than three in the bull are in $S$.
\end{definition}
In Figure~\ref{fig:L-representation}, $\{b,c,4,2,3\}$ is an
$S$-bull but $\{a,b,c,5,6\}$ is not an $S$-bull even though it is
a bull. 

Note: in the remainder of this paper, for a graph $G$ with an 
$[\La]$-representation $R$, we will use $P_x$ to denote the grid path of 
the vertex $x$ of $G$ in $R$.

\begin{observation}\label{obs:gem-on-crown}
Let $G$ be a split graph  with a split partition $(C,S)$.
If $G$ admits an $[\La]$-representation and contains a gem, then
exactly one of the gem's $S$-vertices has its $\La$-path lying
on the crown of the representation.
\end{observation}
\begin{myproof}
Let the vertices of the gem be $c_1$, $c_2$, $c_3$, $s_1$, $s_2$ with $c_1$,
$c_2$, $c_3$ $\in C$, $s_1$, $s_2$ $\in S$ and $s_1 c_1$, $s_1 c_2$, $s_2 c_2$,
$s_2 c_3$ $\in E(G)$. Assume that neither $P_{s_1}$ nor $P_{s_2}$ lies on the
crown. Since $s_1$ and $s_2$ are incomparable, by
Observation~\ref{obs:comparable}, we may assume $P_{s_2}$ lies on
a branch. Since $s_1$ is adjacent to $c_2$, $P_{s_1}$ must lie on
the vertical segment of $P_{c_2}$ and lie above $P_{c_2}$ in the
representation. By our assumption, $P_{s_1}$ must lie on the trunk.
By Observation~\ref{obs:comparable}, $s_1$ dominates $s_2$, a
contradiction. Thus, we may assume $P_{s_1}$ lies on the crown.
Since $s_1$ is incomparable with $s_2$, $P_{s_2}$ cannot lie on the
crown.
\end{myproof}
\begin{observation}\label{obs:bull-on-vertical}
Let $G$ be a split graph  with a split partition $(C,S)$.
If $G$ admits an $[\La]$-representation and contains an $S$-bull,
then some $S$-vertices of this bull have their paths lying on
either the crown or trunk of the representation. \proofBox
\end{observation}
\begin{observation}\label{obs:universal}
Let $G$ be a split graph  with a split partition $(C,S)$.
Suppose there is a vertex $v$ in $G$ with $N(v) = C-\{v\}$. Then
$G$ is an $[\La]$-graph if and only if $G-v$ is. 
\end{observation}
\begin{myproof}
Note that $v$ has no neighbor in $S$.  Suppose $G-v$
has an $[\La]$-representation.  All $\La$-paths of vertices of $C$
contain the base of the crown. We can place $P_v$ so the that it lies at
the top of the base of the crown -- and if necessary, move paths of $S$
on the crown up -- to obtain a representation of $G$. Note that
no $S$-vertices were placed on the trunk since $v$ is inserted 
between the base of the crown and the crown without its base. 
\end{myproof}

%First assume $v \in S$. On
%the trunk, there is a vertical segment where all of $C$ meets. We
%can place $P_v$ there to get a representation of $G$. Now, we may
%assume $v \in C$. Thus, $v$ has no neighbor in $S$. We can place
%$P_v$ on the grid edge at the base of the crown --and if necessary
%move the other paths of $S$ on the crown up-- to obtain a
%representation for $G$.
%\end{myproof}
%
\noindent \textbf{Remark:} ``Moving an $\La$-path up" in an $[\La]$-representation
may require inserting a row into the grid since $\La$-paths start and end
at vertices of the grid.
\begin{observation}\label{obs:twin}
Let $G$ be a split graph  with a split partition $(C,S)$.
Suppose $G$ contains  twins $a$ and $b$. Then $G$ is an $[\La]$-graph
if and only if $G-a$ is. 
\end{observation}
\begin{myproof}
Suppose $a$ is adjacent to $b$. Suppose further that $a$ is in $S$.  Then $b$
is in $C$ and it follows that $a$ is adjacent to all vertices of
$C$. So, we are done by Observation~\ref{obs:universal}. Thus,
we can assume that  both $a$ and $b$ are in $C$. Consider an $[\La]$-representation of
$G-a$. By making $P_a$ an exact copy of $P_b$, we obtain a
representation for $G$.

Now assume $a$ is not adjacent to  $b$. Suppose both $a$
and $b$ are in $S$. Consider an $[\La]$-representation of $G-a$.
Then $P_b$ lies on a branch, on the trunk, or on the crown.
We can assume $P_b$ does not lie on both a branch and the crown or trunk
by moving it up if necessary.  By placing $P_a$
so it lies next to $P_b$ on the branch (or on the trunk, or on the crown, respectively)
that $P_b$ lies on, so that $P_a$ intersects the same $\La$-paths that $P_b$ does,
we obtain a representation for $G$.
%(see Figure~\ref{fig:L-representation} for an illustration.)
Now, we may assume $a$ is in $C$ and $b$ is in
$S$. It follows that $a$ has no neighbor in $S$. But then we are
done by Observation~\ref{obs:universal}.
\end{myproof}
\begin{observation}\label{obs:threshold}
Let $G$ be a split graph  with a split partition $(C,S)$.
Suppose there is a subset $D$ of $C$ such that the vertices of
$X=N(D) \cap S$ are pairwise comparable and $N(X) \subseteq D$.
Then $G$ is an $[\La]$-graph if and only if $G-(D \cup X)$ is.
Further, $G$ can be constructed from $G-(D \cup X)$ so that no $X$-vertex
is placed on the trunk.
\end{observation}
\begin{myproof}
Suppose
there is an $[\La]$-representation of $G-(D \cup X)$. Vertices of
$D$ will be represented by $\La$-paths starting with the base $f$
of the crown so that they all have the same bend-point,
just below the first (highest) branch.
Recall that the $\La$-paths of $C-D$ all contain the base $f$ of the crown.
We can move the $\La$-paths of the $S$-vertices which lie on the crown up so they
do not intersect with the vertical parts of the paths of $D$. We can place the
paths corresponding to vertices  of $X$ so that they lie on this new branch (and
thus not on the trunk).
\end{myproof}

\begin{observation}\label{obs:neighbor-degree-one}
Let $G$ be a split graph  with a split partition $(C,S)$. Suppose
some vertex $c \in C$ is such that all of its neighbors in $S$
have degree one. Then $G$ is an $[\La]$-graph if and only if $G-c$ is. \proofBox
\end{observation}
\begin{observation}\label{obs:gem-comparable}
Let $G$ be a gem-free graph with a split partition $(C,S)$. Then
any two vertices of $S$ with a common neighbor in $C$ are
comparable. \proofBox
\end{observation}
\begin{observation}\label{obs:gem-domination}
Let $G$ be a gem-free graph with a split partition $(C,S)$. Let
$s$ be a maximal vertex in $S$ and $s'$ be a vertex in $S$ with a
common neighbor with $s$. Then $s$ dominates $s'$. 
\proofBox
\end{observation}

\begin{figure}
\centering \hfill
\begin{tabular}{c}
\includegraphics[scale=1]{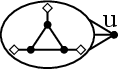} \\ $U_1$
\end{tabular}
\hfill
\begin{tabular}{c}
\includegraphics[scale=1]{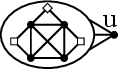} \\ $U_2$
\end{tabular}
\hfill \
\par
\begin{tabular}{c}
\includegraphics[scale=1]{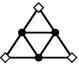} \\ $G_1$
\end{tabular}
\hfill
\begin{tabular}{c}
\includegraphics[scale=1]{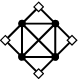} \\ $G_2$
\end{tabular}
\hfill
\begin{tabular}{c}
\includegraphics[scale=1]{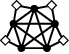} \\ $G_3$
\end{tabular}
\hfill
\begin{tabular}{c}
\includegraphics[scale=1]{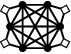} \\ $G_4$
\end{tabular}
\hfill
\begin{tabular}{c}
\includegraphics[scale=1]{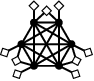} \\ $G_5$
\end{tabular}
\hfill
\begin{tabular}{c}
\includegraphics[scale=1]{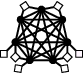} \\ $G_6$
\end{tabular}
\hfill
\begin{tabular}{c}
\includegraphics[scale=1]{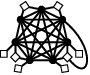} \\ $G_7$
\end{tabular}
\caption{In $U_1$ and $U_2$, the vertex $u$ is adjacent to all
remaining vertices.}\label{fig:characterization}
\end{figure}

Consider the nine graphs shown in
Figure~\ref{fig:characterization}. We believe that they are the
only minimal forbidden obstructions for a split graph to be an
$[\La]$-graph. We pose this as a conjecture.

\begin{conjecture}
A split graph is an $[\La]$-graph if and only if it does not contain any of
the nine graphs in Figure~\ref{fig:characterization} as an induced
subgraph.
\end{conjecture}
Theorems~\ref{thm:bull-free}~and~\ref{thm:gem-free} (proved in the
next sections) can be seen as first steps in this direction.

\begin{lemma}\label{lem:9graphs}
None of the nine graphs shown in Figure~\ref{fig:characterization}
is an $[\La]$-graph.
\end{lemma}
\begin{myproof}
By Lemma~\ref{lem:AT}, the graphs $U_1$ and $U_2$ do not
admit $[\La]$-representations.

Consider the graph $G_1$ with the
split partition $(C,S)$ where $C=\{c_1, c_2, c_3\}$, $S=\{s_1,
s_2, s_3 \}$, and $s_i c_i, s_i c_{i+1} \in E(G)$ with the
subscripts taken modulo 3. Each pair of $S$-vertices is in a gem.
Observation~\ref{obs:gem-on-crown} says that in an $[\La]$-representation
of a gem, exactly one of its two $S$-vertices lies on the crown.  This
is not possible. So, $G_1$ is not an $[\La]$-graph. Similarly,
Observations~\ref{obs:gem-on-crown} and
\ref{obs:comparable} show that $G_2$, $G_3$, and
$G_4$ are not $[\La]$-graphs.

Consider the graph $G_5$. Suppose
$G_5$ admits an $[\La]$-representation. Let $B_1, B_2, B_3$ be the
three $S$-bulls of $G_5$. By
Observation~\ref{obs:bull-on-vertical}, each $B_i$ contains an
$S$-vertex $s_i$ such that $P_{s_i}$ lies on the trunk or crown.
Without loss of generality, we may assume the trunk contains $s_1$
and $s_2$. The fact that $s_1$ is incomparable with $s_2$
contradicts Observation~\ref{obs:comparable}. Similar arguments
show that $G_6$ and $G_7$ are not $[\La]$-graphs.

Finally, it is a
routine but tedious matter to show that all proper induced
subgraphs of the graphs in Figure~\ref{fig:characterization} are
$[\La]$-graphs.
\end{myproof}
%
%THIS WAS MOVED UP.
%We believe the nine graphs in Figure~\ref{fig:characterization}
%are the only minimal forbidden subgraphs for {\em Split} $\cap
%[\La]$. We pose that as a conjecture.
%\begin{conjecture}
%A split graph is an $[\La]$-graph if and only if it does not contain any of
%the nine graphs in Figure~\ref{fig:characterization} as an induced
%subgraph.
%\end{conjecture}
%Theorems~\ref{thm:bull-free}~and~\ref{thm:gem-free} (proved in the
%next sections) can be seen as first steps in this direction.

The {\em k-sun} ($k \geq 3$) is the graph obtained by taking a
cycle on $2k$ vertices and joining every pair of odd-indexed vertices by an edge.
So, a 3-sun is the graph $G_1$, a 4-sun is the graph $G_2$, and
$G_3$ occurs in any $k$-sun with $k \geq 5$. A graph is {\em
strongly chordal} if it is chordal and contains no $k$-sun. The
following follows from Lemma~\ref{lem:9graphs}.
\begin{observation}
Chordal $\cap$ $[\La]$ = Strongly Chordal $\cap [\La]$. \proofBox
\end{observation}
\subsection{Split graphs without S-bulls}
In this section, we give a characterization by forbidden induced
subgraphs of split $[\La]$-graphs without $S$-bulls. This provides
a polytime algorithm for recognizing split $[\La]$-graphs without
$S$-bulls.
\begin{observation}\label{obs:bull}
Let $x_1, x_2$ be two incomparable vertices in $S$. If $G$ does
not contain an $S$-bull, then no vertex $s \in S$ is adjacent to
some vertex $x$ of $N(x_1) - N(x_2)$ and to some vertex $y$ of
$N(x_2) - N(x_1)$.
\end{observation}
\begin{myproof}
If such a vertex $s$ exists, then $\{s, x, y,
x_1, x_2\}$ induces a $S$-bull.
\end{myproof}
\begin{theorem}\label{thm:S-bull-gem-free}
All $S$-bull-free, gem-free split graphs are $[\La]$-graphs.
\end{theorem}
\begin{myproof}
By induction on the number of vertices. Let $G$ be a graph with a
split partition $(C,S)$ and with no $S$-bull. Let $s_1$ be a maximal $S$-vertex.
If $N(s_1) = C$, we are done by Observation~\ref{obs:universal}. So
assume $N(s_1) \neq C$. Let $S_1$ be the set of $S$-vertices which have a neighbor
in common with $s_1$. Since $G$ is gem-free, by Observation~\ref{obs:gem-domination},
$s_1$ dominates all vertices in $S_1$. So $N(s_1) = N(S_1)$. Since $G$ is $S$-bull
free, by Observation~\ref{obs:bull}, the vertices of $S_1$ are pairwise-comparable.
Let $C_1 = N(S_1)$. Then $N(C_1) \cap S = S_1 \cup \{s_1\}$, since if vertex $c \in C_1$
has a neighbor $s \in S$, then since $c \in N(s_1)$, vertices $s_1$ and $s$ have
common neighbor $c$, so $s \in S_1$. By the induction hypothesis,
$G-(C_1 \cup S_1)$ is an $[\La]$-graph.  By Observation~\ref{obs:threshold}, $G$
is an $[\La]$-graph.
\end{myproof}

\begin{theorem}\label{thm:bull-free}
Let $G$ be a graph with a split partition $(C,S)$ and with no
$S$-bull. Then $G$ admits an $[\La]$-representation if and only if $G$ does
not contain $U_1$ or $G_4$ as an induced subgraph.
\end{theorem}
\begin{myproof}
%{\em Sketch of proof.}
By induction on the number of vertices. We
only need to prove the ``if'' part. Let $G$ be a graph with a
split partition $(C,S)$ and with no $S$-bull, $U_1$, or $G_4$.
If $G$ has no gem, the result follows from Theorem~\ref{thm:S-bull-gem-free}.
So we assume that $G$ contains a gem; that is,
there are two incomparable $S$-vertices with a common
neighbor. Let $s_1, s_2 \in S$ be two incomparable $S$-vertices with a
common neighbor such that $d(s_1) + d(s_2)$ is largest, where
$d(x)$ denotes the degree of vertex $x$. Define $C_0 = N(s_1) \cup N(s_2)$.
The following two facts are easy to establish.
\begin{equation}\begin{minipage}{0.8\linewidth}\label{e:domination}
Let $s_3$ be an  $S$-vertex with a neighbor in $C_0$. Then $s_3$ is
comparable to $s_1$ or to $s_2$.
\end{minipage}\end{equation}
Suppose $s_3$ is incomparable
to both $s_1$ and $s_2$. Vertex $s_3$ has no neighbors in $N(s_1)
\cap N(s_2)$, for otherwise it can be shown that $G$ contains
an $S$-bull or $U_1$. Without loss of generality, we may assume
$s_3$ has a neighbor $x$ in $N(s_1) - N(s_2)$. Now, there is a
$S$-bull with vertices $s_1,s_2,s_3,x$, and some $y \in N(s_1)
\cap N(s_2)$. We have established (\ref{e:domination}).

\begin{equation}\begin{minipage}{0.8\linewidth}\label{e:neighborhood-inclusion}
For any vertex $s_3 \in S$ with a neighbor in $C_0$, either $s_1$
or $s_2$ dominates $s_3$.
\end{minipage}\end{equation}
Consider a vertex
$s_3 \in S$ with a neighbor in $C_0$. Suppose $s_3$ has a neighbor
$y \not\in C_0$. By (\ref{e:domination}), we may assume $s_3$ is
comparable to $s_2$. The existence of $y$ implies $s_3$ dominates
$s_2$. It follows that $s_3$ is comparable to $s_1$, for
otherwise, $d(s_3) + d(s_1)
> d(s_2) + d(s_1)$, contradicting our choice of $s_1$ and $s_2$.
Thus, $s_3$ dominates $s_1$. By Observation~\ref{obs:bull}, with
$x_1 = s_1, x_2 = s_2$ and $s =  s_3$, $G$ contains an $S$-bull, a
contradiction. So, we have $N(s_3) \subseteq C_0$. By
Observation~\ref{obs:bull}, $s_3$ has either no neighbor in $N(s_1)-
N(s_2)$ or no neighbor in $N(s_2)- N(s_1)$. Thus,
(\ref{e:neighborhood-inclusion}) is established.

The paths of $ N(C_0) \cap S$ will lie on the
crown and first branch. The paths of $S- N(C_0)$ will lie on
branches below that first branch. By
(\ref{e:neighborhood-inclusion}), the vertices of  $N(C_0) \cap S$
can be partitioned into two sets $D_1$ and $ D_2$ such that $s_i$
is in $D_i$ and dominates every vertex in $D_i - s_i$. Now, we
claim that
\begin{equation}\begin{minipage}{0.8\linewidth}\label{e:Di-comparable}
The vertices in each $D_i$ are pairwise comparable.
\end{minipage}\end{equation}
If some two vertices $x_1, x_2 \in D_i$ are incomparable, then by
Observation~\ref{obs:bull}, $G$ contains an $S$-bull. So,
(\ref{e:Di-comparable}) holds.

It follows that the vertices of
$C_0$ are pairwise comparable in the subgraph of $G$ induced by
 $C_0 \cup D_1$ (and in the subgraph of $G$ induced by $C_0 \cup D _2$).
Vertices of $C_0$ will be
represented by $\La$-paths with the same bend-point. Place the
paths representing $D_1$ so they lie on the crown with $P_x$ being above $P_y$
if $x$ is dominated by $y$. (If two vertices dominate each other,
place one so that it lies above the other.) Place the paths representing $D_2$
so they lie on
the first branch with $P_x$ to the right of $P_y$ if $x$ is
dominated by $y$. (If two vertices dominate each other,
place one so that it lies to the right of the other.) For any two vertices $a,b$ of $D_1$ (respectively,
$D_2$), if $a$ dominates $b$ in $D_1$(respectively, $D_2$), then every
$\La$-path of a $C$-vertex must pass through an edge of $P_a$ to
reach $P_b$. This completes the description of the representation
of $C_0 \cup (N(C_0) \cap S)$.

%Note that if a vertex in $D_i$ has a neighbor in $N(s_1) - N(s_2)$
%(or, $N(s_2) - N(s_1)$), then it is adjacent to all vertices of
%$N(s_1) \cap N(s_2)$ by Observation~\ref:obs:bull}.

Define $C' = C - C_0$. By (\ref{e:neighborhood-inclusion}), there
is no vertex in $S$ with a neighbor in $C'$ and one in $C_0$. The
set $C' \cup (N(C') \cap S)$ contains no gem, for otherwise, $G$
contains $G_4$.
It follows from Observations~\ref{obs:gem-comparable}
and~\ref{obs:gem-domination} that the set $C'$ can be partitioned
into sets $C_1, C_2, \ldots, C_k$ ($k \geq 1$) such that, for each
$i$, the vertices in $N(C_i) \cap S$ are pairwise comparable, and
no $S$-vertex has a neighbor in $C_i$ and one in $C_j$, for $i
\not= j$ (in particular, for each $C_i$, there is a maximal
$S$-vertex $s$ with $N(s) \cap C = C_i$). Define $X= N(C_1) \cap
S$. By the induction hypothesis, $G- (C_1 \cup X)$ is an
$[\La]$-graph. By Observation~\ref{obs:threshold}, $G$ is  an
$[\La]$-graph. 
\end{myproof}

We note that a polytime algorithm to construct an
$[\La]$-representation for the input graph can be extracted from
the proofs above. The algorithm is certifying in the sense that it
produces either an $[\La]$-representation, or an obstruction.

\subsection{Split graphs without gems}
In this section, we give a characterization by forbidden induced
subgraphs of split $[\La]$-graphs without gems. This provides a
polytime algorithm for recognizing split $[\La]$-graphs without
gems. First, we need to introduce a definition.

\begin{figure}[h]
\centering
\includegraphics[scale=1]{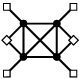}
\caption{The graph $G_8$. Note: this graph contains two disjoint $S$-bulls.}
\label{fig:G8}
\end{figure}

%WRONG CONCEPT
%Two
%vertex-disjoint $S$-bulls are {\em incomparable} if every
%$S$-vertex in one bull is incomparable with every $S$-vertex in
%the other bull.
%
\begin{lemma}\label{lem:two-bull}
Let $G$ be a gem-free graph with a split partition $(C,S)$.
Suppose $G$ does not contain the graph $G_8$ of Figure \ref{fig:G8}
as an induced subgraph. Then,
there is an $[\La]$-representation of $G$ with no $S$-vertices
having their $\La$-paths lying on the trunk. 
\end{lemma}
\begin{myproof}
By induction on the number of vertices.

We can assume that no vertex $v$ has $N(v)=C-\{v\}$ by applying induction and
Observation~\ref{obs:universal}.

It follows that every vertex $c \in C$ has a neighbor in $S$ and that
if $s \in S$ then $C-N(s)\neq \emptyset$. We can also assume that no
$S$-vertex is isolated.
%Kathie forgets why she wanted "no $S$-vertex is isolated."

We partition the vertices of $S$ in the following way.  Let $s_1$ be a
maximal $S$-vertex, let $S_1$ be the set of $S$-vertices with a neighbor
in common with $s_1$, and let $C_1=N(s_1)$. In general, for each $i$, $i \ge 2$,
let $s_i$ be a maximal vertex in $S-(S_1 \cup \dots \cup S_{i-1})$,
let $S_i$ be the set of $S$-vertices which have a neighbor in common with $s_i$,
and let $C_i=N(s_i), (1 \le i \le k)$. By definition of $S_i$, for each i,
$N(C_i)  \cap S = S_i$. Since $G$ is gem-free, by Observation~\ref{obs:gem-domination},
for each $i$, $s_i$ dominates all vertices in $S_i$, so $N(S_i)=N(s_i) = C_i$.
Since $N(s_1) \neq C_1$, there are at least two sets $S_i$ (that is, $k \ge 2)$.
Since $G$ does not contain $G_8$, at least one of the subgraphs induced by
$C_i \cup S_i$  (say, $C_j \cup S_j$), does not contain an $S$-bull. Then by
Observation~\ref{obs:bull}, the vertices of $S_j$ are pairwise comparable.  By
the induction hypothesis, $G-(C_j \cup S_j)$ admits an $[\La]$-representation
with no paths representing the vertices of $S-S_j$ lying on the trunk.  Then
by Observation~\ref{obs:threshold}, $G$ has an $[\La]$-representation with no paths
representing $S$-vertices lying on the trunk.
\end{myproof}
\begin{theorem}\label{thm:gem-free}
Let $G$ be a gem-free graph with a split partition $(C,S)$. Then
$G$ admits an $[\La]$-representation if and only if $G$ does not contain
$G_5$ as an induced subgraph.
\end{theorem}
\begin{myproof}
By induction on the number of vertices. We
only need to prove the ``if'' part. Let $G$ be a gem-free graph
with a split partition $(C,S)$ and not containing $G_5$.
As in the proof of Lemma~\ref{lem:two-bull}, we can assume that
every vertex $c \in C$ has a neighbor in $S$ and that if $s \in S$,
then $C-N(S) \neq \emptyset$. We can also assume that no
$S$-vertex is isolated.
%Kathie forgets why she wanted "no $S$-vertex is isolated."
Define $s_i, S_i$ and $C_i, (1 \le i \le k)$ as in the proof of Lemma~\ref{lem:two-bull}.
Then for all $i$, $N(C_i) \cap S = S_i$. Since $G$ is gem-free,
$s_i$ dominates all vertices in $S_i$, and so $N(S_i) = C_i$. If
the vertices of some $S_i$ are pairwise comparable, then we are done
by the induction hypothesis and Observation~\ref{obs:threshold}.
Therefore, for each $i$, $S_i$ must contain two incomparable vertices,
that is, the subgraph $G[C_i \cup S_i]$ must contain an $S$-bull.  Since
$G$ does not contain $G_5$, it follows that $k=2$ and also that
$G[C_1 \cup S_1]$ does not contain $G_8$.   By
Lemma~\ref{lem:two-bull}, there is an $[\La]$-representation of
$G_1$ with no vertices of $S_1$ on the trunk. By the induction
hypothesis, the graph $G_2=G-(C_1 \cup S_1)$ has an
$[\La]$-representation. We place the branches of $G_2$ under those
of of $G_1$ and extend the vertical segments of the paths of
$C-C_1$ to the crown of $G_1$. The adjacency of $G$ is preserved
because $G_1$ has no $S$-vertices on the trunk in the
$[\La]$-representation.
\end{myproof}
\section{Concluding Remarks and Open Problems}\label{sec:conclusion}
In this paper, we considered the edge intersection graphs of
$\La$-shaped paths on a grid. We showed that recognizing such
graphs is NP-complete. We considered the open problem of
characterizing chordal $[\La]$-graphs. As first steps in solving
this problem, we found characterizations of split gem-free
$[\La]$-graphs and split $[\La]$-graphs without $S$-bulls (a class
more general than split bull-free $[\La]$-graphs). Our
characterizations imply polytime algorithms for recognizing these
two classes of graphs. We posed a conjecture on the
characterization of split $[\La]$-graphs. This conjecture would
imply a polytime recognition algorithm for split $[\La]$-graphs. The
following open problems related to our works arise: (1) Extending
the observations in Section~\ref{sec:split} to other subclasses of
$B_1$-EPG graphs; (2) Find a polytime algorithm for recognizing {\em Chordal}
$\cap \; [\La]$; (3) Establish NP-completeness of recognizing
$B_k$-EPG graphs for every $k$ at least 2.

\section{Acknowledgements}
This research was supported by the Natural Sciences and Engineering Research Council of Canada. 
Additionally, Steven Chaplick was partially supported by the ESF GraDR EUROGIGA grant as project GACR GIG/11/E023.

\bibliographystyle{acm}
\bibliography{B1-EPG}
\end{document}